\documentclass{article}
\usepackage[legalpaper,margin=1in]{geometry}

\newcommand\snowmass{\begin{center}\rule[-0.2in]{\hsize}{0.01in}\\\rule{\hsize}{0.01in}\\
\vskip 0.1in Submitted to the  Proceedings of the US Community Study\\ 
on the Future of Particle Physics (Snowmass 2021)\\ 
\rule{\hsize}{0.01in}\\\rule[+0.2in]{\hsize}{0.01in} \end{center}}
\usepackage{authblk}
\usepackage{hyperref}
\usepackage{comment}
\hypersetup{colorlinks=true,linkcolor=blue, filecolor=blue, urlcolor=blue, citecolor=blue, linktocpage}

\title{\Huge{CompF5: End User Analysis \\ Topical Group Report}}

\author[1]{Gavin~S.~Davies}
\author[2]{Peter Onyisi}
\author[3]{Amy Roberts}
\affil[1]{\href{mailto:gsdavies@olemiss.edu}{gsdavies@olemiss.edu}; Department of Physics \& Astronomy, University of Mississippi, University, MS 38677, USA}
\affil[2]{\href{mailto:ponyisi@utexas.edu}{ponyisi@utexas.edu}; Department of Physics, University of Texas, Austin, TX 78712 USA}
\affil[3]{\href{mailto:amy.roberts@ucdenver.edu}{amy.roberts@ucdenver.edu}; Department of Physics, University of Colorado Denver, Denver, CO 80217, USA}

\date{(and contributors from the community)}

\begin{document}

  \maketitle
  \snowmass{}







\section{Introduction}
When talking about End User Analysis we refer to the extraction of physics results from reconstructed and simulated experimental data. High energy physics experiments produce systems for performing common reconstruction, calibration, and simulation tasks which result in shared data samples. These detailed data samples are then reduced to produce a range of analysis samples, often optimized for a particular physics topic by trigger or data object selection. End users (analyzers) then analyze those samples to produce physics results.  Hidden information and lack of documentation can make those physics results more difficult to produce and these difficulties are not felt equally; scientists at institutions with limited software expertise, scientists who are uncomfortable pushing for other's time, and scientists who aren't a priority for an expert face unnecessary burdens when parts of the analysis ecosystem rely on expert intervention or training. 

Data analysis is generally I/O intensive and repeated many times during an experiment's lifetime as selections and algorithms are refined. In many cases end user analysis is performed on local computing clusters, or even laptops, although dedicated facilities for parallel analysis of large samples also play an important role.

Analysis activities include, but are not limited to:
\begin{itemize}
  \item Calibration
  \item Feature detection (Higgs discovery)
  \item Limit setting (searches)
  \item Parameter estimation (W mass measurement, CP violation)
  \item Cross section measurements
\end{itemize}

Community discussions converged on categorizing End User Analysis into five distinct sub-topics: \textit{Analysis Ecosystems} which includes Libraries, Languages, and Data Formats, \textit{Analysis Models}, \textit{Dataset Bookkeeping and Formats}, \textit{Collaborative Software}, and \textit{Training}.
This report identifies impediments to end user analysis and potential avenues to address such issues, that have been informed by discussions with the user community and backed up by submitted letters of interest, white papers, and other publicly available material. Community discussions included sessions in the Computational Frontier workshop~\cite{CompF-workshop} on August 10 -11, 2020, the CompF5 Topical Group Town Hall~\cite{CompF5-TownHall} on February 28, 2022 and a community survey.

\subsection{Requirements for a Sustainable End-User Software Ecosystem}
A choice of software stacks is becoming available for HEP analysis. It is generally thought that the ROOT and Python stacks have different strengths and will competitively coexist into the future.

In order for the ecosystems to achieve long-term, sustainable success, we note the following requirements:
\begin{itemize}
    \item \textbf{Support of Personnel.} Unsupported software projects undergo ``code rot'' over time, a process where changes external to the package itself cause it to lose functionality. In the Python ecosystem, for example, the migration from Python 2 to Python 3 rendered old versions of packages unusable. For this reason alone, it is mandatory that any software that forms a key part of a HEP analysis ecosystem must have a maintainer (person or organization) able to provide the necessary level of support for the package. 
    \item \textbf{Maintenance Support} The community must understand that maintenance is a task of equivalent importance to developing new code, and recognize people's work appropriately. There must also be a mechanism for transferring responsibility for a package as necessary.   
    \item \textbf{Documentation and Training.} Analysis software ecosystems are used as a gestalt --- not as a disconnected set of packages. 
    \item \textbf{Interoperability.} Ecosystem lock-in can be a problem for multiple reasons. Packages that can only be used in a single ecosystem will not provide benefits to users not in that ecosystem. Interoperability is critical for innovation that has broad impact.
\end{itemize}

The HEP Software Foundation aims to play a key role in organizing the long-term evolution of (among other things) HEP analysis ecosystems, and has produced summaries and roadmaps with similar information and conclusions as this report \cite{HSF,Pivarski:2022ycs}. The IRIS-HEP project \cite{IRISHEPWEB}, a NSF Software Institute, is a model of a large-scale effort distributed over many stakeholders to develop tools to address the computing needs of the High Luminosity LHC; this may potentially be replicated in other frontiers, or in a cross-frontier effort.



\section{Analysis Ecosystems: Libraries, Languages, and Data Formats}
No analysis software functions entirely on its own; any package is situated in the context of the input data it consumes, the output data it produces, the other software it depends on, and the way it is configured or embedded in other code. Because of this we talk about ``software ecosystems,'' groups of packages which are typically used together.

Packages in an ecosystem typically have common data interchange formats and similar programming language interfaces. In some cases they may be distributed together as a single metapackage with overall versioning. There are two major ecosystems in HEP:
\begin{itemize}
    \item \textbf{ROOT} \cite{Brun:1997pa}: the ROOT suite is a tightly-integrated set of libraries that cover a broad range of HEP analysis needs, including I/O, event loop execution (including a parallel distributed mode), histogramming, fitting and statistical analysis, and visualization. The libraries are written in C++ and that is still considered the primary language for its use, although the Python bindings (PyROOT) are very well supported and by construction expose essentially the full API.  Bindings for the R language are also currently supported. The ROOT libraries were developed to meet the specific needs of HEP experiments and as such provide solutions that are well-matched to HEP analysis problems, although this also means that use outside of HEP is limited. The integrated, tightly-bound nature of ROOT means that using alternative software for any particular functionality can be difficult. ROOT is undergoing a redesign \cite{Naumann:2022pub} (``ROOT 7'') which aims to improve interfaces with the benefit of modern C++ and the benefit of over 25 years of practical experience. ROOT is hosted by CERN with significant contributions from US partners, particularly in the area of I/O.
    \item \textbf{Python} \cite{van1995python,10.5555/1593511}: this is a somewhat loose term for a set of tools, with Python as the primary language interface, introduced with the primary goal of enabling the use of software developed outside of HEP, in particular for machine learning. Many core packages in this ecosystem are part of the Scikit-HEP~\cite{Rodrigues:2020syo} and the Institute for Research and Innovation in Software in HEP (IRIS-HEP)~\cite{IRISHEPWEB} projects. This ecosystem is still in development and has no single governance team; the IRIS-HEP and HEP Software Foundation have invested heavily in Python libraries \cite{Pivarski:2022ycs}. It tends to emphasize independent packages for different aspects of the analysis pipeline (so, for example, I/O is handled with a different package from histogramming). Due to a standard and extensive set of tools aimed at supporting development of open source Python packages (the pypi package repository, github actions for continuous integration, documentation hosting on readthedocs) the barriers to entry for new software in this ecosystem are quite low. Where packages connect with well-supported libraries, scientists can enjoy documentation and support of a community much larger than HEP.  Development teams in this ecosystem are typically small and feature junior personnel.
\end{itemize}

The long history of ROOT in HEP means that similar featuresets have sometimes been developed more than once, in incompatible fashions. For example, parallel processing of events on a single node was supported by PROOF-Lite \cite{Ganis:2008zz}, the implicit multithreading feature for executing queries on TTrees, and RDataFrame \cite{Piparo:2019xdy}. In the Python ecosystem a similar situation arises due to simultaneous development and shorter development cycles.

Ecosystems associated with other programming languages, notably Java \cite{Chekanov:2020bja}, Go \cite{Binet:2018xcc}, and Julia~\cite{bezanson2017julia,Stanitzki:2020bnx,Gal:2022vnn}, have been developed and seen use in certain situations. Adoption of these is coupled to use of the relevant languages in the community, which at the moment is minimal.

\subsection{Programming Languages}
Users interact with software libraries and packages through programming languages. These can be separated into general-purpose languages (GPLs) which can be used for any task, and domain-specific languages (DSLs) which provide a restricted set of higher-level primitives which simplify certain operations. The two most commonly-used general-purpose languages in HEP are C++ and Python. Examples of domain-specific languages are the TCut syntax used for applying selections and constructing new variables in ROOT, and the kumac language used to control the old FORTRAN-based PAW suite.

General-purpose languages, by definition, are extremely capable and are used to solve problems outside of HEP. Exposure to these languages is one of the major technical skills that is transferable outside the field. It is considered necessary for HEP students to develop familiarity with, and preferably proficiency in, at least one general-purpose language. It is not necessarily the case that the languages that are used in HEP are the ones prevalent in the industries that particle physicists transition to (for example, R is widely-used in data science and virtually unknown in HEP). Because of their complexity, the time it takes to train personnel in them, and the need to transfer responsibilities for maintaining code from one person to another, the diversity of general-purpose languages used in the field is strictly limited. By contrast, domain-specific languages historically have been easier to master due to the limited range of constructs available.

HEP has had relatively few general-purpose languages in recent history. Until the early 2000s FORTRAN was commonplace. A desire to move to more modern and commonly-used languages drove a transition across the field to C++ (although there was competition, notably from Java); this was generally a top-down move imposed as new experiments were built or experimental upgrades were implemented. Often experiments found it valuable to use a second general-purpose language such as Python or tcl as a high-level scripting system for their data processing code. Python in particular came to be adopted by users as a convenient language and its simultaneous adoption as the standard language in machine learning has driven massive bottom-up adoption of the language. Availability of library bindings in various languages is extremely important for their use; both C++ and Python raise significant barriers to using libraries written in those languages elsewhere, although thanks to a lot of work the border between those two specific languages is relatively low.

Python is not necessarily an optimal language for scientific computing~\cite{876288}. In its reference implementation, it is a fully interpreted language, making it much slower than C++ for many tasks. The speed issue creates a programming paradigm in which users express operations via intermediate libraries (such as numpy~\cite{harris2020array} for data manipulation or TensorFlow~\cite{tensorflow_developers_2022_6574269} for neural network construction), introducing what amount to mini-languages embedded in Python. For this reason there is interest in exploring languages that can combine the expressiveness and ease-of-use of Python with compilation to machine code; the most commonly-explored option is Julia. Introducing another general-purpose language in HEP will require a very compelling case and it appears that the status quo regarding Python will continue for the foreseeable future, perhaps including the adoption of acceleration technologies such as numba~\cite{10.1145/2833157.2833162}.

Visualizations are increasingly being steered from web browsers (such as Jupyter notebooks~\cite{soton403913} or JSROOT \cite{JSROOT}). By far the dominant language in that environment is JavaScript, which is a language which particle physicists generally have very little experience in, and one where best practices have evolved very rapidly. If critical parts of the analysis ecosystem are written in this language, expertise will need to be maintained at some level.

Ideally it would be possible to seamlessly combine code written in different languages in a single application. This is extremely difficult to achieve in the general case due to languages' different memory layouts, calling conventions, and assumed invariants. Specific pairings, such as Python/C++, have solutions for interoperation. A general solution could involve leveraging languages' foreign function interfaces (FFI) through some common layer, such as the LLVM \cite{LLVM:CGO04} Intermediate Representation.

Domain-specific languages in HEP span a range of applicability, including steering reconstruction workflows \cite{Bennett:2022gyi}, specifying operations in columnar analysis \cite{Proffitt:2021wfh}, specifying likelihood construction from histograms \cite{TRexFitter}, and describing analysis at a very high level \cite{Prosper:2022lnf}. In a meaningful sense this includes the expressions needed to control numpy~\cite{harris2020array}/AwkwardArray~\cite{jim_pivarski_2020_3952674}/TensorFlow~\cite{tensorflow_developers_2022_6574269} from Python, or to express operations using the ROOT RDataFrame \cite{Piparo:2019xdy}; although formally these are library operations, analysts are expected to learn how to express their intent in terms of high-level operations while the details of execution are kept intentionally opaque. 

Domain-specific languages have the advantage of providing high-level primitives which in principle permit optimization of the actual execution of the code. In particular this includes parallelization and acceleration, operations which analysts may be uncomfortable with or which require significant investment to implement properly but which can provide speed and capability improvements when available. The major disadvantage is the (usually) restricted scope of operations that can be expressed in DSLs, which are typically designed with specific tasks in mind and which may make it unnatural or impossible for users to do other things. This is particularly dangerous for DSLs that operate at the ``analysis description'' level; it is not clear that such languages can generalize reasonably between frontiers.

Users can come to regard DSLs as primary parts of the analysis interface and the learning curve for them can be high. In particular, if there is more than one DSL relevant to a specific task, users may prefer to learn only one. If DSLs are linked to specific libraries or ecosystems, the synergies are liable to tie users to those technologies.

\subsection{Data Formats}
Analysis data come in many forms:
\begin{itemize}
\item \textbf{Event data:} these consist of information, typically with a fixed but complex schema, describing individual events.
\item \textbf{Histograms:} These summarize features extracted from event data.
\item \textbf{Other summary data:} There are forms of summary data that cannot reasonably be expressed via histograms: sometimes tabular data is a better fit and more space-efficient, and sometimes the schema for the data is sufficiently complex that it makes sense to store a sui generis kind of object (such as for the results of fits).
\item \textbf{Configuration data:} The configuration for running some software may be stored in a human-readable and -editable format, or in a binary format --- the latter is especially common when one package is configuring the operation of another. In either case, interacting with the stored configuration requires data access, and it may be possible to alter the configuration like any other kind of data. 
\item \textbf{Metadata:} For end-user analysis, this tends to primarily be provenance tracking and version information.
\end{itemize}

File formats can describe both the overall container for data and the specific types of objects that can be stored. ROOT separates these fairly strictly, in that the ROOT file format can store essentially any C++ object, and ROOT objects can be serialized to formats other than ROOT files (such as JSON or XML). ROOT provides a number of pre-defined data objects, such as tables (known as TTrees) and histograms, and multiple objects can be present in the same file. Other data formats allow less freedom; for example, Apache Parquet \cite{Parquet} merges the container and the data object and is only suitable for tabular data, while HDF5 \cite{HDF5} files natively allow for a specific set of contained structures.

It is important to note that analysis end-users very rarely interact directly with underlying file formats --- they work with the in-memory transient representation of data, rather than the persistent format, and the translation between the two is handled by libraries. The capabilities of specific formats may limit what users can do, and certain formats may provide more optimized storage, but otherwise the details are generally hidden from users. Therefore transitions in data format are easier to handle than those in libraries or languages. Newer versions of ROOT include the capability to read in data in CSV text format, sqlite files, or Apache Arrow~\cite{Arrow}.

The structure of read requests to persistent storage, and the translation between on-disk and in-memory formats (decompression and deserialization), can significantly affect the speed of data analysis. ROOT is working towards the introduction of a replacement for the TTree class called RNTuple \cite{Blomer:2020usr} which in some cases can increase data read speed by over an order of magnitude compared to TTrees \cite{Lopez-Gomez:2022umr}. The HDF5 format, by comparison, suffers from a lack of multi-threaded I/O and shows poor performance compared to RNTuple and Apache Parquet. Data storage format choices in the future will need to ensure that the format itself does not constitute a bottleneck for analysis speed.

\subsection{Visualization}

End-user analysis relies critically on visualization to give feedback to the physicist. Plots allow the user to quickly understand characteristics of the data but also to debug code and workflow problems.

Previous generations of analysis libraries supported interactive visualization through native graphics libraries; remote use involved the use of technologies such as remote X Windows. Over high latency links this can be extremely difficult to use and requires specific software to be installed on the user's machine. Recent trends exploit the near-universal availability of web browsers following common standards to offload rendering and interaction to the user's browser. This results in a more uniform experience across platforms and reduces external dependencies. This is the standard mode for code in Jupyter~\cite{soton403913} notebook environments (in particular the Python ecosystem) and is the baseline for future ROOT graphics.

The ROOT ecosystem's visualization libraries are naturally matched to the specific HEP use case. Visualization in the Python ecosystem is typically handled through the Python matplotlib~\cite{Hunter:2007} package, a standard for scientific plot creation; matplotlib does not directly support a number of common HEP plot forms or histogram structure formats, so additional libraries have been written to help bridge these boundaries \cite{mplhep, hist}.

Although event displays are not typically used as part of an analysis workflow, they are still of interest to end users, and the same issues apply. Work is being done on experiment-agnostic event displays that render in browsers using JavaScript \cite{Phoenix}.

\subsection{Findings and Recommendations}
\textbf{Finding:} There are two primary ecosystems developing for analysis in the next decade: one based on the ROOT libraries and one based on a collection of Python libraries. These projects have similar goals and scope but are organized differently and have different philosophies regarding industry tool use.\\
\textbf{Recommendation:} Development of both ecosystems should be supported. The friendly competition between the two has already resulted in significant improvements for users. Maintaining interoperability between the two (e.g. in data formats) should be a priority.\\

\textbf{Finding:} The main general purpose languages used in HEP today are C++ and Python. Projects such as PyROOT have enabled interoperability of these languages at a level not replicated by other options in HEP.\\
\textbf{Recommendation:} These two languages both have important roles to play in their respective niches and are expected to remain dominant in the near future. Other languages are generally unfamiliar to the community, have weak interoperability with existing libraries, and impose a maintenance burden if used without careful planning. Outside of bottom-up projects in other languages, or overwhelming domain-specific needs (e.g. JavaScript for visualization in web browsers), C++ and Python should be recommended. There is a reasonable concern about the efficiency of using interpreted languages in core analysis kernels, both in terms of computing resource use and consequent environmental impact. We recommend that the Python ecosystem community demonstrate that this issue has been addressed, and also ensure that just-in-time compiler technology (e.g. numba) is easy to integrate into user code.\\

\textbf{Finding:} A number of domain-specific languages have been proposed to address problems in various spheres. Additionally, certain uses of Python (e.g. when scripting ML libraries, or working with RDataFrame) can be viewed as a DSL as they require library-specific knowledge.\\
\textbf{Recommendation:} ``High-level'' DSLs that attempt to describe analysis via high-level objects are unlikely to generalize well between experiments as often the ``primitives'' are too different, but can be useful in certain situations. The lower-level DSLs can be extremely useful where relevant, however effort should be taken to avoid unnecessary duplication of scope as this imposes burdens on users similar to using multiple general purpose languages.\\

\textbf{Finding:} The ROOT container file format is ubiquitous in HEP, and comes with a serialization scheme tightly linked to the ROOT libraries. Industry and non-HEP tools typically use other formats (e.g. Apache Parquet for the ROOT TTree), and efforts are ongoing to enable their use. In addition the ROOT team foresees an evolution of the TTree to the more-optimized RNtuple. Multiple independent implementations of ROOT I/O are available. As ROOT is specialized to the HEP use case, it supports features missing from other formats.\\
\textbf{Recommendation:} The ROOT file format is extremely important for ongoing experiments and historical data and compatibility must be maintained. Other formats have important use-cases. Tools to translate between formats, and to enable various ecosystems to ingest and produce them, should be maintained.\\

\textbf{Finding:} Software typically has low barriers to initial entry but significant ongoing maintenance requirements. New software projects are frequently initiated without much concern for long-term support (and this is reasonable since many bottom-up projects do not succeed).\\
\textbf{Recommendation:} Computing personnel should be funded specifically to maintain software projects identified as satisfying an important need in the community (for example, by the HEP Software Foundation~\cite{HSF}).

\section{Analysis Models}
Once analysis code is written, it must be run on data. An analysis pipeline may involve multiple stages of data reduction and different codes, executing on very different platforms.

Analysis users value fast turn around --- being able to quickly answer physics questions. Because the rapidity with which analysis workflows can complete is paramount, data processing architectures which are well-suited to managed production workflows may not match well on to analysis tasks. Users often desire to run analyses on hardware that is under their control, and in fact such capability may be essential to allow users to develop, test, and debug their code. 

\subsection{Scale}
Users need to be able to scale code execution from a few events (for testing) to an experiment's full dataset (for actual analysis). The former requires interactive response, while the latter may require distributed execution (on a single cluster or across multiple sites). Interactive use historically has occurred via terminal sessions and visualization software native to the particular operating system and environment. Distributed execution has occurred on batch systems, either single-site or multi-site; in the latter case, especially if a federated computing model is adopted, complex issues of data locality and access, bookkeeping, job brokering, fair access, and so on arise.

Particle physics problems are usually \textit{high-throughput}, not \textit{high-performance}, problems: that is, they consist of a very large number of fairly lightweight computations which are essentially independent of each other, and so do not require computing resources to appear as a single, very powerful image (as on a traditional supercomputer). Traditional tightly-coupled supercomputer execution environments such as OpenMPI~\cite{gabriel04:_open_mpi} are therefore not typically needed for HEP analysis applications and in fact may be detrimental as they do not exploit the fine granularity of HEP problems. However, the increasing exploitation of coprocessors and accelerators (such as GPUs) in HEP code requires libraries that couple CPU and GPU execution on a single node.

Adequate hardware resources need to be provided to users. From the user perspective the main feature of the hardware side of the execution environment is how it is provisioned (latency to obtain resources, whether the resources are heterogeneous, how explicit the user needs to be about the scale of needed resources). New developments in \textit{analysis facilities} \cite{Benjamin:2022dpo,Adamec:2021vkl} leveraging technologies from industry (such as Apache Spark \cite{Spark} and Kubernetes \cite{Kubernetes}) hold the potential to provide users with responsive, autoscaling access to compute.

Software solutions that smoothly scale from small-scale interactive tests to full-data processing are desirable. In the absence of such solutions users face a barrier during the development of their analyses which can be quite substantial. Such solutions need to abstract away the execution of the event loop so that it can be executed on whatever resources are available, transparently to the user. Therefore by necessity they restrict the form of the user code to some extent. Packages that attempt to provide such transparent scaling include Coffea \cite{Smith:2020pxs} in the Python ecosystem and Distributed RDataFrame \cite{Padulano:2020wmf} in the ROOT ecosystem.

\subsection{Interfaces}

The ``traditional'' HEP analysis execution environment is a terminal. GUI applications add significantly more coding complexity and generally limited benefits, although they have been used (for example, in the ROOT PROOF suite, which still generally is invoked via a terminal). The use of a terminal allows terminal scripting languages, such as bash, to be used to orchestrate a workflow.

There is a recent trend towards using \textit{notebooks}, particularly those provided by the Jupyter environment, as the main interface for user code execution. Jupyter notebooks, which are rendered in a browser but with a connection to a backend kernel at the actual code execution site, function essentially as recorded interactive terminal sessions with additional documentation and output visualization capabilities. Jupyter notebooks are ill-suited for execution as actual code and are instead primarily used to script libraries which (for example) spawn worker tasks to actually perform the requested computations.  Despite these limitations, Jupyter notebooks have proven increasingly popular with early-career scientists, highlighting the need for access to an environment that requires little to no installation or setup; part of the appeal of Jupyter notebooks is that they can be hosted from a cluster and provide a containerized environment that is already set up for collaboration analysis.  Jupyter notebooks also provide a way to interactively execute code and display plots, which can otherwise be difficult when running analysis at facilities that forbid X-forwarding for security reasons.

\subsection{Findings and Recommendations}
\textbf{Finding:} Users need to be able to scale their analyses from simple tests on a small debugging dataset to full deployment over all data. In many cases this transition requires working in a different environment (a common case is transitioning from a local workstation, referring to specific files, to a batch job running on a catalogued dataset). This can involve significant difficulties for users and cause support issues.\\
\textbf{Recommendation:} There is unlikely to be a one-size-fits-all solution for all experiments. Recent work with interactive analysis facilities which provide extreme scaling capabilities to users through a single interface may address many of these issues; if successful the resulting software stacks should be made available to small experiments in a turnkey way.\\

\section{Dataset Bookkeeping and Formats}

Data in the Computational Frontier falls into four categories, and each has an array of associated formats.  ROOT files are inescapable and present in virtually every type of analysis, although there are other formats in use, often to support efforts like machine-learning where libraries outside the ROOT ecosystem have developed rapidly:   

\begin{itemize}
  \item Raw data: overwhelmingly, custom binary formats that are specific to an experiment.  Interest in accessing these files to explore new reconstruction methods is growing.
  \item Processed and/or skimmed data: predominantly ROOT.
  \item Intermediate files that facilitate tool use: Parquet, Zarr, etc.  HDF5 is rarely used, its libraries are a bit painful and all the tools people want to use work with file formats that have friendlier IO libraries.
  \item Metadata (file metadata, run information, processing states): Databases, stand-alone text files, wikis.  Metadata is also often included in the data file.
\end{itemize}

The challenges we face that are directly related to file formats are

\begin{itemize}
  \item \textbf{Preservation.}  Reproducibility, training, open science, and principles of Findable, Accessible, Interoperable, and Reusable (FAIR)~\cite{Wilkinson2016} are increasingly of interest and importance to the community.  Making full use of our data requires documentation about the format, software that can read the format and provide analysis utility, and extensive metadata so that analyzers can make interpret or make use of the data.  Related efforts include TOPS, and more.  There are significant international investments in these efforts; US-based efforts will require funding models beyond the traditional short-term grant cycle.  Preservation efforts must be guaranteed for at least 10 years to see community uptake.
  \item \textbf{Interoperability.} Re-using already-developed tools is an appealing way to reduce the software development burden that slows down experiments across the field.  To do this, data formats must be compatible with these tools.  The desire for interoperability has driven experiments to consider non-ROOT file formats like Parquet and HDF5.  This need has also supported tools like uproot, which can read in ROOT files and present them in data formats that are compatible with common Python libraries.
\end{itemize}

\subsection{Data Cataloging and Delivery}

Users need to be able to obtain the input data for their analyses at the sites where their computation occurs. This means they, or the analysis infrastructure, must be able to discover where the data are stored and handle any data transfer/networking issues involved in bringing the data and compute resources together.

Data catalogs are useful even when the data are all stored in a single site, since they group related files into datasets and store useful, queryable metadata. When storage resources are distributed then catalogs become even more critical. Data placement is beyond the scope of this report but can be handled by systems that also provide a data catalog, such as Rucio \cite{Barisits:2019fyl}.

Data delivery to analysis tasks can be handled in a number of ways. On a small scale, the data may be provided via a POSIX filesystem (possibly networked). At a large scale, the data may need to be copied or streamed between locations over a wide area network. When multiple sites are involved, issues of authentication and identity management need to be addressed; maintaining multiple distinct accounts can be a heavy burden to users. Federated identity management and token-based authorization can address these issues.  Data transfer to a user job from remote storage can be handled in various ways; a common HEP standard is the xrootd protocol \cite{xrootd}, which robustly supports partial file reads (useful for minimizing network bandwidth if only a subset of data are to be read). Large-scale, distributed storage in industry is generally provided via object stores, for which the S3 protocol is the natural choice.

Standalone files that are considered opaque to data-handling and cataloging software, and only interpreted by end-user code, are not the only possible way to store and provide data. ``Intelligent'' infrastructure can provide a database-like view of datasets, enabling event selection and column reduction as a service (see e.g.\ \cite{Galewsky:2020xig}). User code need never interact with the ``native'' data format at all. In such an architecture the backend data store may itself be structured as a database \cite{Gutsche:2020kmd}. Work in this direction may be able to leverage advances in computational storage technology.

\subsection{Findings and Recommendations}
\textbf{Finding:} Experiments usually have global solutions for handling metadata involved in dataset cataloging and workflow control. However these solutions frequently are not available for individual analyses and users may need to develop bespoke solutions.\\
Small experiments often struggle to implement even the global solutions; implementation difficulty does not scale down with experiment size and smaller experiments often lack personnel to devote to metadata systems.\\
\textbf{Recommendation:} Effort should be put into developing user-friendly data provenance and metadata storage systems that can be easily integrated into typical analysis tasks.  Support for software developers across collaborations and support for multi-experiment metadata systems could help smaller experiments.

\section{Collaborative Software}

At its heart, HEP research is collaborative in nature and as the complexity of the physics problems increases this necessitates the need to effectively and efficiently communicate and collaborate with peers.
The size and available expertise of an experiment may dictate what tools can be employed. When talking about collaborative software one is considering code management and distribution, version control, communication software, forums, wikis, and Q\&A platforms.
\subsection{Tools for Collaboration}

\begin{itemize} 
  \item Code Management and Version Control\\
In terms of code management and version control, many people love git~\cite{chacon2014pro}. Equally many feel that more git training is needed. However, some experiments also support subversion~\cite{subversion} and CVS (the Concurrent Versioning System), in some cases for legacy code, and so expertise and exposure is still necessitated
within experiments for these additional alternatives. Effort should be made to establish the use of a version control platform at the early stage of a user's research career. It is especially important to note that version control is the basis for many important workflows, including:
  \begin{itemize}
    \item testing via continuous integration for reproducibility and quality control
    \item automated packaging and installation
    \item open review and transparency
    \item documentation of software changes and issue tracking       
  \end{itemize}
  \item Communication\\
In order to stay informed and in contact with collaborators all users use some form of instant messaging platform for both synchronous and asynchronous communication. Video and voice conferencing is dominated by the uptake of Zoom but other formats exist and are still used, such as Skype, Vidyo. Email and phone communication are fast being overtaken as the primary method of communication by several platforms that offer IRC-style features, persistent channels (or chat 'rooms') organized by topic, private groups, and simple person-to-person direct messaging. Examples of such messaging programs include Slack, Zulip, Mattermost, Discord, Discourse. 
Many of these offerings provide integration with various communication tools to be able to seamlessly switch between them (for example, Slack integration with GitHub, Zoom, and Google tools) but often at a premium, per-user cost basis (albeit with education/research based cost reduction). Indeed programs such as Zulip are open-source and afford more control and zero financial burden at the cost of humanpower and resources to maintain them within a collaboration.
  \item Software distribution\\
Once software is built, mechanisms are needed to deliver the software to end users. The advent of the CernVM-File System (CernVM-FS) has greatly enhanced software distribution throughout the HEP field providing a scalable, reliable and low maintenance service~\cite{HEPSoftwareFoundation:2020daq,Blomer:2011zz,7310920}. The long-term maintenance and development of this important infrastructure needs to be accepted as new challenges arise, notably scaling issues. \\
External software products are managed and installed on file distribution services (i.e. CernVM-FS) through various different mechanisms. An example, for Fermilab-based experiments in particular, is access to external
products provided by a Fermilab-developed product-management package called Unix Product
Support (UPS), for selecting particular build flavors, and Unix Product Distribution (UPD) for uploading/downloading of products between local systems and distributions servers. Alternatives emerging and being adopted within HEP include (but not limited to) open source package managers Spack~\cite{7832814}, Anaconda's conda~\cite{anaconda}, and the package installer for Python, pip~\cite{pypi}. All of these options still require the necessary humanpower and expertise to deploy and maintain. 
The proliferation of different packaging options is not a new problem and was studied in detail within a HEP Software Foundation Packaging Working group report~\cite{l_sexton_kennedy_2016_1472340}. Such fruitful, frequent discussions and summaries are critical in order to keep abreast of recent developments, understand common problems with the aim of converging on common solutions. 

Containers are now also commonplace components for job submission. Some pieces of software are so platform dependent that it is sometimes nearly impossible to replicate the framework on another machine, thus containers are a popular way to deploy applications but they do also come with some disadvantages, such as whilst they are a good tool for job submission, graphical applications do not work well, persistent storage is complicated,...for example. It is also another tool that users must be familiar with in their workflow and should have appropriate documentation and training materials.
  \item Project Management\\
Fewer users employ bug-tracking or project management tools. Some users use commercial programs such as Trello or Notion for project management on the individual level (far too expensive for large experiments) or simple, free Google Docs/Sheets. There is some migration to using GitHub Issues.
  \item Documentation\\
Any given experiment may have found a good solution for distributing software, or the right tool to facilitate communication amongst users but without appropriate attention to documentation this can lead to communication gaps, uninformed decisions, lack of transparency, wasted time, inefficient onboarding of new users, and an increased knowledge gap between experts and users.  Good documentation tools and practices can alleviate some of the pain of knowledge transfer, in the case of experts retiring or moving on to new projects/fields and new users joining. In practice a number of different tools are employed to encourage documentation: technical notes, wikis, Doxygen, Sphinx, GitHub wiki/issue tracking to name a few. Even with the right tool in place, the encouragement and incentive needs to be in place to provide said documentation and keep it up to date. This also couples with the training discussion in Sec.~\ref{sec:training}. Internal experiment-specific surveys identify the need for improved documentation and working examples as key to end users being able to perform analyses.

\end{itemize}

\subsection{Accessibility}
The topic of accessibility for end users overlaps with training \& documentation. It is difficult to quantify how much of “how to do things” lives in private or hard-to-search communication forums/channels
How much do these tools enable “chatting with colleagues” and can we improve that?

Experiment analysis software \& environments should be encouraged to guide people towards constructing reproducible and archivable analyses from the start.

\subsection{Findings and Recommendations}
\textbf{Finding:} The “full” analysis stack of an experiment also includes software that enables interaction between analyzers. This includes documentation of the experiment's code, messaging between users, discussion forums, software version control, bug tracking, and document workflow management. Especially for small experiments, setting up this infrastructure from scratch can be daunting both in effort and cost.\\
\textbf{Recommendation:} Access to a full stack of these services should be provided to funded experiments.  XSEDE/ACCESS could also play an important role here as some laboratories have onerous access requirements.

\section{Training}\label{sec:training}

Computational and software training needs are widespread in our field; we discuss these more broadly than just analysis software because even as an "end user" interacting only with data, understanding details of multiple systems like the triggering, data acquisition, and simulation are almost always necessary to do science.  In this section we attempt to summarize the training recommendations --- for software alone --- that appeared in over 100 unique white papers submitted to SNOWMASS: we do not have the expertise we need for the computational work across the Cosmic Frontier.  The question our community must address is: where can this expertise come from?  Listed below are the current training efforts common in the computational frontier. 

\begin{description}
  \item [Training through physics curriculum] Departments often offer and sometimes require a Computational Physics course.  PICUP~\cite{Caballero2019-qj} is an organization focused on integrating computing with the undergraduate physics education.  However, most students arrive at their research groups and require weeks to years of software training before being able to do science.
  \item [Training within groups] burdens groups without access to computational experts and disadvantages students who are uncomfortable or discouraged from asking questions.  And it is rare for a single group to have a research software engineer who might act as a mentor; research software engineers are often supported on soft money and computing work is not typically valued in the promotion or tenure process.
  \item [Training within collaborations] is ideally a strong incentive for reproducible analyses that can serve to train incoming scientists.  Larger-scale collaborations like LSST have had success supporting quality training, but smaller collaborations often struggle to train the software experts they need.
  \item [Training within fields] FIRST-HEP is a leading effort in this space and is aiming to make commonly-needed, Carpentry-style lessons for HEP-specific computing skills.  DANCE-Edu is a similar effort, focused on the Dark Matter and Neutrino communities.  There are many topics that are common across the field, and high-quality curriculum that is findable and available year-round is a critically needed.  Training within the Computational Frontier, or at least across similar experiments, is the focus of this section as the overlap is strong enough that material needs minimal to no change to be useful to someone else.
  \item [Training across fields] Biology, Geology, and other data-intensive fields have some skills overlap with the Computational Frontier and this broad need is part of the success of the Carpentries.  Training across fields is beyond the scope of this report.
  \item [Leveraging training from other disciplines] is most evident in Electric Engineers, who are often partners on low-level data acquisition needs.  In general the need for these skills far outstrips personnel and asking individuals to add training and mentoring to their design workload is not effective.  Computer science seems like it should be another area of training overlap, but they have a higher pay scale and much of our work is not interesting from a computer science perspective. It is crucial that we formulate our domain's problems (in the form of compute, productivity, and I/O challenges for example) often and in great detail to computer scientists if we hope to benefit from their expertise.
  \item [Industry training partnerships] Successful collaborations do exist such as the highly beneficial engagement of NVIDIA with domain scientists for over a decade.  However, most students who need introductory training do not receive it through industry partnerships or through professional training programs.   
\end{description}

Sustainable efforts in HEP computation require continual recruitment and training of  the HEP workforce and the rest of this section focuses on field-wide training efforts. The HEP experimental community has realized the importance of training and tries to meet this need through experiment-specific training and, increasingly, through the use of common training materials and training instruction, notably The Carpentries project~\cite{Carpentries} and HEP specific produced by the HEP Software Foundation~\cite{Malik_2021}.  Notably, The Carpentries provides both training material and has a procedure in place for training their trainers.

Training needs range from on-boarding of novice summer students to perform a simple Python coding task, through experimental training weeks for graduate students and postdocs, to high-level training of domain experts in specialized subjects and training for the trainers themselves. Practitioners vary from people running prepackaged analysis code on well defined data samples to experts writing machine learning algorithms for field-programmable gate arrays (FPGAs) or designing multi-national workflows for the WLCG.  

A well trained HEP computing workforce has enormous payoffs, by avoiding wasted effort and resources, by optimizing the quality and reproducibility of physics results, by providing our colleagues with marketable skills, and sharing advanced expertise in computing with the institutions we work for.  In addition to their research, HEP faculty (and our astrophysics colleagues) are often the driving force behind the introduction of  modern computing into the undergraduate Physics curriculum and thus have influence far beyond our field.  

Many of the training materials we currently use are one-off talks given by experts within the context of a single experimental training effort.  But many topics are general across multiple efforts.  The field would benefit greatly by building on the growing efforts to identify common learning modules,  to curate existing tools when available and, if not, to write them using best pedagogical practices and make certain they are widely available and supported.    Such common training brings in wider effort and encourages the use of common tools.  A recent example is the cross-experiment Dark Matter and Neutrino Computation Explored (DANCE)~\cite{Roberts:2022ezy} program offered in conjunction  with Computational and Data Science Training for High Energy Physics (CoDaS-HEP)~\cite{CODAS-HEP} at  the Snowmass Community Summer Study workshop.

In the past, such common training efforts have suffered from the  tight connection between research funding and specific experiments, which discouraged cross-experimental efforts. US funding agencies are now making substantial efforts in support of common computing efforts, including training and documentation, notably through the NSF-funded IRIS-HEP program, other NSF Cybertraining proposals, and DOE's recent call for graduate training grants proposals. These efforts, in collaboration with colleagues worldwide, should be encouraged and extended.

\subsection{Findings and Recommendations}
\textbf{Finding:} There are many cases where experiments are limited by the availability of personnel trained in various aspects of computing. All students need to be trained in the software and languages used for analysis; beyond that, specialized training may be needed for specific areas, such as GPU use or code optimization. Many traditional avenues (e.g. dedicated academic classes or industry training) may not be a good match to needs or affordable. There are significant efforts through the HEP Software Foundation and other initiatives to develop effective training for HEP software. Software training has a direct connection to concerns about diversity, equity, and inclusion.\\
\textbf{Recommendation:}  We need to support collaborations focused on learning best practices from education research, industry, and assessment. In addition, we need to explore and understand the scalability of various approaches to training within our field.  Funding should be made available for the development and hosting of training materials, instructor training, and should include partnerships with education researchers to ensure these efforts are effective.

\section{Long-Term Preservation}
The scope and details about Long-Term Preservation is covered by the dedicated \textit{Reinterpretation and Long-Term Preservation of Data and Code} topical group~\cite{Bailey:2022pdq} within the Computational Frontier therefore we direct the reader to that section of the full report. Nevertheless we provide a recommendation in the context of end users and their interaction with long-term preservation.
\subsection{Findings and Recommendations}
\textbf{Finding:} Transitioning from the way an analysis is actually done to a “packaged” version that can be rerun by an outsider from scratch is typically complex, as the entire workflow is rarely described in a single place. This causes people to see long-term preservation as an additional burden whose benefits they will not see.\\
\textbf{Recommendation:} Provide pipelines to nudge users into choosing practices compatible with long-term preservation as the default. These need to be considered and built in to the structure of analysis systems at the start, not bolted on at the end.

\section{End Users}

At the core of end user analysis are the users themselves; the people doing the software work. Before users can consider pursuing innovative ideas, the general challenge of maintaining existing software and computing infrastructure persists. This is a problem exacerbated further in small experiments~\cite{FASER:2022yqp}. The use of common tools can partly mitigate this problem but this does not address the lack of long-term support for software efforts, when one considers a typical grant funding cycle is three years. Incentives are needed for supporting existing code versus re-inventing the wheel when unnecessary, although well-motivated innovation has and will continue to be critical to improving the end-user experience and science output. One concern is that letting "a thousand flowers bloom" can also lead to a long-term maintenance problem; we need pathways for maintenance of well-supported libraries and state-of-the-art work should consider how continuation of their projects will align with the preservation of core libraries. 

The field must find ways to recognize sustainable software work and support stable career paths for scientists interested in this critical work. There is all too often a misalignment between what the field recognizes as valid work and what experiments need in terms of functioning software. In the case of early career scientists, software maintenance and documentation can often sideline them for the traditionally recognised physics analysis work. Additionally, software efforts are siloed by collaboration; it is rare for users to be funded for cross-experiment software efforts. Development of tools for one particular experiment can often times find equal utility for other experiments~\cite{Backhouse:2022cgc}. We need mechanisms for people to obtain partial support to work beyond their immediate use cases. We must have support mechanisms for both bottom-up and top-down development \& long-term support.

\subsection{Findings and Recommendations}
\textbf{Finding:} A lot of transformative ideas are introduced to the HEP computing community and are implemented by early-career scientists (especially grad students and postdocs). Support for these physicists can be minimal, and the career trajectories for scientists interested in software work are limited.\\
\textbf{Recommendation:} Software work (especially with cross-experiment application) should receive stronger consideration for funding. More cross-experiment/frontier computing physicist positions could be created.  Funding agencies and frontiers need to work together to identify viable long-term funding patterns for this work.
\clearpage




\bibliographystyle{JHEP}
\bibliography{myreferences}




\end{document}